\documentclass[4p]{elsarticle}

\usepackage{amssymb}

\journal{Nuclear Physics A}

\begin{document}

\begin{frontmatter}



\title{From Kuo-Brown to today's realistic shell-model calculations}


\author[label1]{L. Coraggio}
\author[label2]{A. Covello}
\author[label1]{A. Gargano}
\author[label1,label2]{N. Itaco}
\address[label1]{Istituto Nazionale di Fisica Nucleare, \\
Complesso Universitario di Monte  S. Angelo, Via Cintia - I-80126 Napoli,
Italy}
\address[label2]{Dipartimento di Fisica, Universit\`a
di Napoli Federico II, \\
Complesso Universitario di Monte  S. Angelo, Via Cintia - I-80126 Napoli,
Italy}

\begin{abstract}
This paper is an homage to the seminal work of Gerry Brown
and Tom Kuo, where shell model calculations were performed for
$^{18}$O and $^{18}$F using an effective interaction derived from the
Hamada-Johnston nucleon-nucleon potential.
That work has been the first successful attempt to provide a 
description of nuclear structure properties starting from the free
nucleon-nucleon potential.
We shall compare the approach employed in the 1966 paper with the
derivation of a modern realistic shell-model interaction for
$sd$-shell nuclei, evidencing the progress that has been achieved
during the last decades.
\end{abstract}


\end{frontmatter}


\section{Introduction}\label{intro}
The paper by Tom Kuo and  Gerry Brown (KB) \cite{Kuo66} is a true
milestone in the theory of nuclear structure.
It has indeed been the first  attempt to perform a microscopic
nuclear-structure calculation starting from the  free nucleon-nucleon
($NN$) potential  which resulted  in a quantitative description of the
spectroscopy  of nuclei belonging to the $sd$ shell.

The KB  work was grounded in the general belief, which came out
between the end of 1950s and the beginning of 1960s, that a new
generation of nuclear structure calculations for finite and infinite
systems, based on first principles and free from phenomenological
inputs, had to be started.
As a matter of fact, $NN$ potentials such as the Yale \cite{Lassila62}
and the Hamada-Johnston (HJ) ones \cite{Hamada62} were able to fit
reasonably well the two-nucleon scattering data, both potentials having
an infinite short-range repulsion and the one-pion-exchange tail.
The handling of the hard-core component of the $NN$ potential in
many-body systems was studied by Brueckner and coworkers
\cite{Brueckner54,Brueckner55}, who introduced an effective potential
- the well-known reaction matrix $G$ - which overcomes the singularity
at short distances via an infinite sum of particle-particle ladder diagrams.
Soon after the Brueckner work, several shell-model calculations were
performed,  where the effective interaction was taken to be the $G$
matrix  (see, for instance, ~\cite{Kahana69} and references therein).

A main step forward was then made by Bertsch \cite{Bertsch65}, who 
studied the role played by the core-polarization diagram
corresponding to one-particle-one-hole ($1p-1h$) excitations at second
order in perturbation theory.
This work  evidenced that this diagram, dubbed ``bubble'', was
responsible for a correction to the interaction as large as $30\%$ of
the first-order contribution, when considering the $^{18}$O and
$^{42}$Sc nuclei and using as interaction vertices the $G$-matrix
elements derived from the Kallio-Kolltveit potential \cite{Kallio64}. 

Brown and Kuo were well aware that the time was ripe to assemble the
new tools, and drew a red line that starting from the free $NN$
interaction ended to the spectroscopic description of a many-nucleon
system within a sound theoretical framework.
The $sd$-shell effective interaction of \cite{Kuo66} was derived
starting from the HJ potential, whose hard-core component was
renormalized via the calculation of the reaction-matrix $G$.
The latter was then employed in the interaction vertices of the
perturbative expansion of the effective hamiltonian, and this
expansion was performed including  terms up to second order.

The KB  shell-model effective interaction obtained within this
approach was used to calculate the energy spectra of $^{18}$O and
$^{18}$F yielding  good agreement with experiment.  
This paved the way to a wide sequence of studies, dedicated to both
the developement of the perturbative approach to the derivation of the
shell-model effective interaction and  the assessment of its role in
the study of nuclear structure (see for example
\cite{Kuo90,Hjorth95,Coraggio09a} and references therein). 

In this paper, we compare the results obtained in 1966  for $sd$-shell
nuclei with those achievable by using a moderm shell-model effective
hamiltonian starting from a high-precision $NN$  potential.
The aim is to give an idea of the progress made along the line traced
by  Kuo and Brown more than 50 years ago.

As is well known, during the last decade $NN$ potentials derived within
the chiral perturbation theory (ChPT) have provided an approach to the
problem of nuclear forces that is well grounded in the quantum
cromodynamics.
The original idea of deriving realistic two- and three-nucleon forces
(2NF and 3NF) within the framework of the effective field theory dates
back to Weinberg \cite{Weinberg79,Weinberg90,Weinberg91}, who
considered the most general Lagrangian involving pions and low-energy
nucleons consistent with the spontaneously broken chiral symmetry. 
The short-range  parts of the potential  are given in terms of
low-energy constants fitted to two-nucleon and, possibly,
three-nucleon data.
 
There are two main advantages in ChPT: the first one is that it
generates nuclear two- and many-body forces on  equal footing
\cite{Wei92,Kol94,ME11}; the second one is that the $NN$ potentials
derived within such a framework  are naturally tailored for the
low-energy regime of the nuclear structure physics. 
This may  allow to avoid the complications of  renormalizing the
short-range repulsion.

We have constructed a shell-model effective hamiltonian $H_{\rm eff}$
starting from the so-called N$^3$LOW nucleon-nucleon potential
\cite{Coraggio07b}, a low-momentum potential derived from ChPT at
next-to-next-to-next-to-leading order with a sharp momentum cutoff at
2.1 fm$^{-1}$.
The theoretical single-particle (SP) energies and the two-body matrix
elements (TBME) of the effective interaction are then obtained within
the framework of the time-dependent degenerate linked-diagram
perturbation theory \cite{Kuo90}, which is an extension of the
approach followed in the KB paper.

In the next section we give a few details about the perturbative
expansion of the effective shell-model hamiltonian, together with a
sketch of the approach followed in the KB work.
In Section \ref{results} the results obtained for $^{17,18}$O and
$^{18}$F with the KB hamiltonian and with that derived from the
N$^3$LOW potential will be reported and compared with the experimental
spectra.
Comments and conclusions are drawn in Section \ref{conclusions}.

\section{Theoretical framework}\label{theo}
As mentioned in the Introduction, in Ref. \cite{Kuo66} the first step
in the derivation of the effective interaction was to overcome the
difficulty of the short-range singularity due to the hard-core of the
$NN$ potential.
This problem was tackled  calculating the Brueckner reaction matrix
$G$ from the HJ potential with the tools available at that time.

As a matter of fact, the calculation of the $G$-matrix was splitted
into two sub-problems; when dealing with the attractive components of
the potential, more precisely the singlet- and triplet-even channels,
the $G$-matrix was calculated using the  Moszkowski-Scott separation
method \cite{Moszkowski60}, which is essentially based on dividing the
potential into a short-range part $V_s$ and a long-range one $V_l$.

The separation method cannot be employed when dealing with the
components that are repulsive outside the hard-core region, since the
essence of this method is that the attractive $V_s$ has to balance the
short-range repulsion.
This is not the case of the singlet- and triplet-odd components of the
HJ potential, for which the calculation of the $G$-matrix was carried
out using the reference-spectrum method \cite{Bethe63}.

It is worth recalling that  during the 1970s more advanced techniques
for the calculation of  the $G$ matrix were developed. 
Suffice it to mention here the method proposed by Tsai and
Kuo~\cite{Tsai72}, which allows a practically exact calculation of
the $G$ matrix and  has been largely employed until the early 2000s.    
In these years, however, a quite new approach to the renormalization
of  the $NN$ potential was proposed \cite{Bogner02}, which consists in
constructing a low-momentum $NN$ potential, $V_{\rm low-k}$, that
preserves the physics of  $V_{\rm NN}$ up to a cutoff  $\Lambda$. 
This approach has proved  to be an advantageous alternative to the
Brueckner G-matrix method and  has become by now a main tool to handle
the short-range repulsion of $NN$ potentials like, for instance, those
based on the meson theory of nuclear forces.  
We ourselves have routinely used this approach in several realistic
shell-model calculations \cite{Coraggio09b,Coraggio13b} employing the
CD-Bonn $NN$ potential  \cite{Machleidt01b}.

In the calculations performed in the present work, however,  there was
no need to renormalize the $NN$ potential. 
In fact, the $N^3$LOW potential, while reproducing accurately the
experimental deuteron binding energy,  low-energy scattering
parameters, and  phase-shifts of $NN$ scattering up to at least 200
MeV laboratory energy, is a smooth interaction that can be used
directly in the derivation of the shell-model hamiltonian.

Starting from this   potential, we have derived the shell-model
effective hamiltonian within the many-body perturbation theory, as
developed by Kuo and coworkers through the 1970s
\cite{Kuo90,Hjorth95}.
More precisely, we have used the well-known $\hat{Q}$-box plus
folded-diagram method \cite{Kuo71}, where the $\hat{Q}$-box is a
collection of one- and two-body irreducible valence-linked Goldstone
diagrams.
Within this framework the effective  hamiltonian $H_{\rm eff}$ can be
written in an operator form as

\begin{equation}
H_{\rm eff} = \hat{Q} - \hat{Q'} \int \hat{Q} + \hat{Q'} \int \hat{Q} \int
\hat{Q} - \hat{Q'} \int \hat{Q} \int \hat{Q} \int \hat{Q} + ~...~~,
\end{equation}

\noindent
where the integral sign represents a generalized folding operation,
and $\hat{Q'}$ is obtained from $\hat{Q}$ by removing first-order
terms.
In the present calculations the $\hat{Q}$-box includes  all diagrams
up to third order \cite{Coraggio10a,Coraggio12a},  and the
folded-diagram series is summed up to all orders using the Lee-Suzuki
iteration method \cite{Suzuki80}.
We sum over the intermediate states between successive vertices whose
unperturbed excitation energy is less than $E_{max}=16~\hbar \omega$,
which is sufficiently large to ensure that the matrix elements of the
effective hamiltonian are almost independent from the value of $E_{max}$.

\begin{figure}[t]
\begin{center}
\includegraphics[width=6.0cm,angle=0]{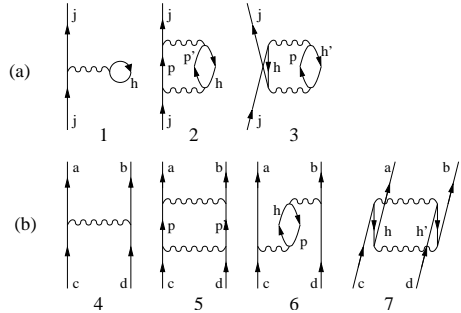}
\end{center}
\caption{\label{figura}
Second-order Goldstone diagrams, with antisymmetrized interaction
vertices, included in the KB perturbative expansion of the shell-model
hamiltonian. (a) labels the collection of the one-body diagrams, (b)
labels the two-body ones. For the sake of simplicity, the ``bubble''
diagrams differing by exchanges of the valence particle labels are not
reported.}
\end{figure}

The  effective hamiltonian  so obtained contains one- and two-body
terms, but we use a subtraction procedure so that only the two-body
term is retained while the SP energies are taken from experiment.
In particular, in the calculations of $^{18}$O and $^{18}$F presented
here we consider $^{16}$O as closed core and take the two SP energy
spacings of the $sd$ space from the experimental spectrum of $^{17}$O
\cite{nndc} while the absolute energies are determined from the
experimental binding energies of $^{17}$O and $^{17}$F with respect to
$^{16}$O \cite{Wang12}, as was done in the KB paper.
The two-body term employed in the latter was instead calculated
considering only corrections to the $G$ matrix as arising from
$2~\hbar \omega$ $1p-1h$ excitations of the $^{16}$O core. 
Within our procedure, this corresponds to include a unique
second-order diagram, the  "bubble'' diagram 6 Fig.~\ref{figura}b, in
the calculation of the $\hat Q$-box, without performing the
folded-diagram expansion. 
It is worth noting that  the latter  is necessary to remove violations
of time ordering coming from the factorization of diagrams from second
order on.  
Actually,  corrections other than the $1p-1h$ ones, corresponding to
diagrams 5 and 7 of Fig. \ref{figura}b, were also discussed in
\cite{Kuo66} but not explicitly included in the calculations.
It is also worth noting that the number of intermediate states taken
into account when calculating the bubble diagram was not enough to
achieve a satisfactory convergence, as was later pointed out in the papers
by Vary et al. \cite{Vary73}, Kung et al. \cite{Kung79}, and Sommermann et
al. \cite{Sommermann81}.

In the following section, the spectra  for $^{18}$O and $^{18}$F
obtained with the KB and $N^3$LOW effective interactions are compared
with experiment.
For the sake of completeness, we also compare  our calculated SP
energies for $^{17}$O,  obtained consistently with the theory
described above,  with those of the KB paper resulting from inclusion
of the three one-body first and second-order diagrams shown in
Fig.~\ref{figura}a.

\section{Results}\label{results}
\begin{figure}[h]
\begin{center}
\includegraphics[width=6.0cm,angle=0]{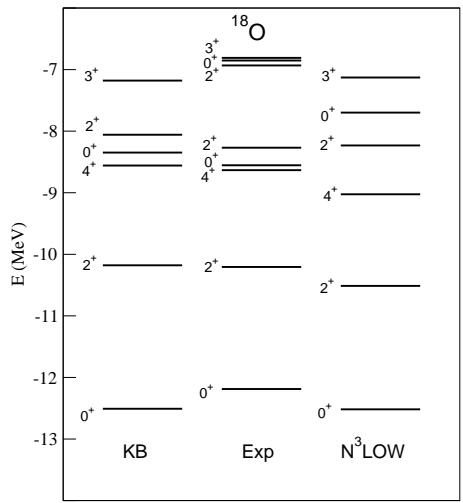}
\end{center}
\caption{\label{18O}
Experimental \cite{nndc} and calculated low-energy spectra for $^{18}$O (see text for details).}
\end{figure}

Let us start with the spectrum of $^{18}$O. 
In Fig. \ref{18O} the experimental low-energy spectrum referred to the
$^{16}$O ground-state energy is compared with those obtained from the
KB and $N^3$LOW effective interactions. 
Note that the KB spectrun is scaled  by twice the  experimental
$^{17}$O ground-state energy with respect to that reported
in~\cite{Kuo66}. 
In the KB work, it is clearly shown that the inclusion of  $1p-1h$
corrections to the $G$ matrix interaction leads to a lowering of the
low-lying states and a raising  of the higher ones, providing a
substantial improvement in the description of  $^{18}$O.  
As a matter of fact, as we see from Fig.~\ref{18O}, the agreement of
the  KB results with experiment is quite good.  
The binding energy is overestimated by only  about 300 keV, and the
differences between the observed and calculated spacings do not go
beyond 500 keV.  
The $N^3$LOW calculations, when excluding the second $0^+$ state,
lead to a further improvement. 
In particular, the energies of the yrast $2^+$ and $4^+$ states come
very close to the experimental values.

The energy of the $0_{2}^+$ state is largely overestimated by our
calculation. 
Our predicted level lies  about 1.2 MeV above the experimental one,
which may be traced to the fact that this state contain  a relevant
collective $4p-2h$ component as testified by the large experimental
value of the $B(E2;0^+_2 \rightarrow 2^+_1)=17\pm2$ W.u. \cite{nndc}.
It is indeed  suprising that the discrepancy reduces to  700 keV in
the KB case, where no $2p-2h$ corrections to the $G$ matrix were
included.

\begin{figure}[h]
\begin{center}
\includegraphics[width=6.0cm,angle=0]{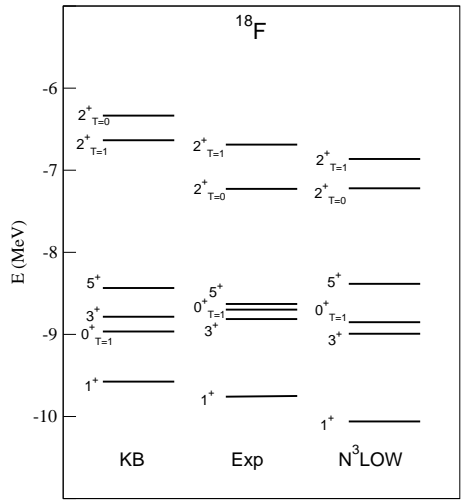}
\end{center}
\caption{\label{18F}
Experimental \cite{nndc} and calculated low-energy spectra for
$^{18}$F (see text for details).}
\end{figure}

We now come to discuss the results for $^{18}$F, which provides a
direct test of the proton-neutron interaction. 
In  Fig. \ref{18F}, the two calculated spectra are compared with the
experimental one.
As for the $^{18}$O case, the KB spectrum shown in the figure is
scaled with respect to the original one by using the experimental
ground-state energies of the two one-valence-particle nuclei.

We see that, although the energy differences between experiment and
theory for both calculations  are quite similar to those found  for
$^{18}$O, the KB spectrum, at variance  with the  N$^3$LOW one, does
not give  the correct sequence of the low-lying $T=0,1$ states.
As concerns the effects of including $2p$ and $2p-2h$ excitations in
the derivation of the effective interaction, no definite conclusions
were drawn in the KB paper owing to possible-double counting or
convergence problems related to the adopted scheme.

Finally,   in Table \ref{tab_17O}  we compare the  SP energy spacings
obtained with KB and N$^3$LOW  interactions with the experimental
ones between the $J^{\pi}=\frac{5}{2}^+,~\frac{1}{2}^+,~\frac{3}{2}^+$
states that have the largest spectroscopic factors in the
$^{16}$O(d,\rm{p)}$^{17}$O transfer reaction \cite{nndc}.

\begin{table}[h]
\begin{center}
\begin{tabular}{| l | l | l | l |}
\hline 
$J^{\pi}$ & KB &  N$^3$LOW & Expt. \\ \hline\hline
$\frac{5}{2}^+$ & 0.0 (-2.64) & 0.0 (-3.211)    & 0.0 (-4.144) \\ 
$\frac{1}{2}^+$ & 0.11          & 0.835            & 0.871    \\ 
$\frac{3}{2}^+$ & 5.51          & 6.281            & 5.085    \\
\hline
\end{tabular}
\end{center}
\caption{Experimental and theoretical SP energy spacings of $^{17}$O
  obtained with KB and N$^3$LOW potentials (see text for
  details). The ground-state energies with respect to $^{16}$O are
  reported in parenthesis.
\label{tab_17O}}
\end{table}

We see that both calculations reproduce reasonably well the observed
SP energy spacings. 
More precisely, the $\epsilon_{1/2^+} - \epsilon_{5/2^+}$ spacing
calculated with N$^3$LOW comes closer to the  experimental value while
the KB interaction yields a spin-orbit splitting $\epsilon_{3/2^+} -
\epsilon_{5/2^+}$ in better agreement with experiment.
The larger value  of the spin-orbit splitting obtained with the
N$^3$LOW calculation can be ascribed to the contribution to the
perturbative expansion coming from third-order diagrams.
As a matter of fact, we have verified that, starting from the N$^3$LOW
potential and using the folded-diagram expansion with a $\hat Q$-box
up to second order, namely by including only the three one-body
diagrams of  Fig.~\ref{figura}a,  we obtain $\epsilon_{1/2^+} -
\epsilon_{5/2^+}=0.626$ MeV and $\epsilon_{3/2^+} -
\epsilon_{5/2^+}=4.945$ MeV,  the latter value being very close to the
KB one shown  in Table~\ref{tab_17O}. 
As regards the binding energies, the values obtained from the
N$^3$LOW and KB calculations are  both within acceptable limits, the
former being only about 900 keV smaller than the experimental value.

\section{Summary and conclusions}\label{conclusions}

In this work, we have revisited the approach to realistic shell-model
calculations followed by Gerry Brown and Tom Kuo in their pioneering
paper \cite{Kuo66}, where the Hamada-Johnston potential was employed
to study the spectroscopic properties of  $^{18}$O and $^{18}$F. 
The KB paper represents a breakthrough in the history of nuclear
structure, since it showed for the first time that it was possible to
reach a reasonable degree of accuracy in the description of
many-nucleon systems starting from the free $NN$ potential.

For the sake of comparison, we have constructed a modern realistic
shell-model hamiltonian and performed calculations for the same nuclei
$^{18}$O and $^{18}$F. 
We have employed  the chiral N$^3$LOW $NN$ potential within the
framework of the $\hat Q$-box plus folded-diagram method, which is
substantially an upgrade of the perturbative expansion carried out in
the KB paper.

As we have already pointed out in Section 2, the shell-model
calculations employing realistic effective interactions have by now
entered the mainstream of nuclear structure theory, having proved to
be able to provide an accurate description of the spectroscopic
properties of nuclei  in different mass regions (see, for instance,
Ref. \cite{Coraggio09a,Coraggio09b,Coraggio14a} and references therein). 
However, in this contribution to the Gerry's memorial volume,  we
deemed  it appropriate to only focus on the two nuclei studied in the
KB paper,  as the starting point of a new generation of nuclear
structure calculations.

From the results presented in Section 3 it appears that the modern
calculations lead on the whole to a moderately  improved description
of the experimental data with respect to the KB results.
This shows the substantial soundness of the original KB approach, which was
able to catch  the main aspects of the physics of many-nucleon systems
within the framework of the shell model.
However,  as we have discussed in Section 2,  the  framework for
realistic shell-model calculations has substantially improved over the
initial one and rests  now on solid theoretical foundations, as
regards both the starting $NN$ potential and the many-body technique
for constructing the effective interaction. 
It is indeed very gratifying that the long journey initiated by Gerry
Brown and Tom Kuo  has been crowned with success.




\section*{References}
\bibliographystyle{elsarticle-num} 
\bibliography{biblio}





\end{document}